\documentstyle[12pt]{article}
\normalbaselines
\newcommand{\be}{\begin{equation}}
\newcommand{\ee}{\end{equation}}
\newcommand{\ba}{\begin{eqnarray}}
\newcommand{\ea}{\end{eqnarray}}
\newcommand{\baa}{\begin{eqnarray*}}
\newcommand{\eaa}{\end{eqnarray*}}
\newcommand{\bb}{\begin{thebibliography}}
\newcommand{\eb}{\end{thebibliography}}
\newcommand{\ci}[1]{\cite{#1}}
\newcommand{\bi}[1]{\bibitem{#1}}
\newcommand{\lab}[1]{\label{#1}}
\newcommand{\re}[1]{(\ref{#1})}

\newcommand\fac[2]{\mbox{$\frac{#1}{#2}$}}

\newcommand\DGM{XX^{th} }

\begin{document}

\begin{flushright}
  UdeM-LPN-TH94-92
\end{flushright}

\vbox {\vspace {10mm}}
\begin{center}

{\large \bf DEFORMATION OF SUPERSYMMETRIC AND \\[2MM]
CONFORMAL QUANTUM MECHANICS  \\[3MM]
THROUGH AFFINE TRANSFORMATIONS}%
\footnote{Talk presented at the Workshop on Harmonic Oscillators,
College Park, 25-28 March 1992}\\

\vspace{10mm}

Vyacheslav Spiridonov%
\footnote{On leave of absence from the Institute for Nuclear Research,
Moscow, Russia}\\
{\it Laboratoire de Physique Nucl\' eaire,
Universit\' e de Montr\' eal, \\
C.P. 6128, succ. A, Montr\' eal, Qu\' ebec, H3C 3J7, Canada}\\[5mm]
\end{center}

\vspace{2mm}

\begin{abstract}
Affine transformations (dilatations and translations) are used to define
a deformation of one-dimensional $N=2$ supersymmetric quantum mechanics.
Resulting physical
systems do not have conserved charges and degeneracies in the spectra.
Instead, superpartner Hamiltonians are $q$-isospectral,
i.e. the spectrum of one can be obtained from another (with possible
exception of the lowest level) by $q^2$-factor scaling.
This construction allows easily to rederive a special self-similar potential
found by Shabat and to show that for the latter a
$q$-deformed harmonic oscillator algebra of Biedenharn and Macfarlane
serves as the spectrum generating algebra. A general class of potentials
related to the quantum conformal algebra $su_q(1,1)$ is described.
Further possibilities for $q$-deformation of known solvable potentials are
outlined.

\end{abstract}

\newpage

\section{Introduction}

Standard Lie theory is known to provide very useful tools for description
of physical systems. Elegant applications were found in quantum mechanics
within the concept of spectrum generating,
or, dynamical (super)symmetry algebras \ci{r1}. The most
famous example is given by the harmonic oscillator problem
(so the name of this workshop) where spectrum is generated by the
Heisenberg-Weyl algebra.
Some time ago a wide attention was drawn to the deformations
of Lie algebras which nowdays are loosely called "quantum algebras", or,
"quantum groups" \ci{r2} (below we do not use the second term because
Hopf algebra structure is not relevant in the present context).
Spin-chain models were found \ci{r3}
where Hamiltonian commutes with generators of the quantum algebra
$su_q(2)$, deformation parameter $q$ being related to a coupling constant.
Thus, an equivalence of a particular perturbation of the
interaction between "particles" to the deformation of symmetry
algebra governing the dynamics was demonstrated.

Biedenharn and Macfarlane introduced $q$-deformed harmonic oscillator
as a building block of the quantum algebras \ci{r4,r5}.
Various applications
of $q$-oscillators appeared since that time [6-13] 
(an overview of the algebraic aspects of $q$-analysis is given
in Ref.\ci{r8}). Physical models refering to $q$-oscillators can be
conditionally
divided into three classes. The first one
is related to systems on lattices \ci{r9}.
In the second class dynamical quantities
are defined on "quantum planes" -- the spaces with non-commutative
coordinates \ci{r10}. Although Schr\" odinger equation in this approach
looks similar to the standard one, all suggested explicit realizations of it in
terms of the normal calculus result in purely finite-difference
equations. Parameter $q$ responsible for the non-commutativity of quantum
space coordinates serves as some non-local scale
on the continuous manifolds and, therefore, the basic physical principles
are drastically changed in this type of deformation.
We shall not pursue here the routes of these two groups of models.

The third -- dynamical symmetry realization class -- is purely
phenomenological: one deforms already known spectra by postulating
the form of a Hamiltonian as some combination of
formal quantum algebra generators \ci{r11}, or, as an anticommutator of
$q$-oscillator creation and annihilation operators \ci{r4,r9}.
This application, in fact, does not have straightforward
physical meaning because of the non-uniqueness of deformation
procedure. Even exact knowledge
of a spectrum is not enough for precise reconstruction of an interaction.
For a given potential with some
number of bound states one can associate another potential
containing new parameters and exhibiting the same spectrum \ci{r12}.
Therefore the physics behind such deformations is not completely
fixed. Moreover, for a rich class of spectral problems there are
powerful restrictions on the asymptotic growth of discrete eigenvalues
\ci{r12a} so that not any ordered set of numbers can represent a spectrum.
All this means that one should more rigorously  define physical interaction
responsible for a prescribed deformation of a given simple spectrum.
$q$-Analogs of the harmonic oscillators were also used for the
description of small violation of
statistics of identical particles \ci{r11c} (general idea on the
treatment of this
problem on the basis of a parametric deformation of commutation relations
was suggested in Ref.\ci{r15}).
The papers listed above represent only a small fraction of works devoted
to quantum algebras and $q$-analysis. For an account of
unmentioned here applications we refer to reviews \ci{r16a,r16b}.

Recently Shabat have found one-dimensional reflectionless potential
showing peculiar  self-similar behavior and describing an infinite
number soliton system \ci{r18}.  Following this development
the author proposed \ci{r17}
to take known exactly solvable Schr\" odinger potentials and try to deform
their shape in such a way that
the problem remains to be exactly solvable but the spectrum acquires
complicated functional character.
So, the Shabat's potential was identified in Ref.\ci{r17} as a
$q$-deformation of conformally
invariant harmonic and particular forms of Rosen-Morse and
P\"oschl-Teller potentials. The hidden $q$-deformed Heisenberg-Weyl algebra
was found to be responsible for purely exponential character of the spectrum.
In comparison with the discussed above third group of models present
approach to "quantum" symmetries is the direct one --
physical interaction is fixed first
and the question on quantum algebra behind prescribed
rule of $q$-deformation is secondary.

In accordance with this guiding principle a
deformation of supersymmetric (SUSY)
quantum mechanics \ci{r19,r20} was proposed in Ref.\ci{r18a}.
This talk is devoted to description of the results of Refs.\ci{r18,r17,r18a}
and subsequent developments. We start by giving in Sect.2 a brief account
of the properties of simplest $(0+1)$-dimensional SUSY models.
In Sect.3 we describe a deformation of these models on the basis of
pure scaling transformation of a superpartner potential, namely, we find
$q$-SUSY algebra following from this rule and analyze its properties.
Sect.4 outlines possible extensions of the simplest  potential
deformation. In Sect.5   we show that
mentioned above self-similar potential naturally appears within
$q$-SUSY as that characterized by the simplest structure of Hamiltonian.
In this case factorization operators
entering the supercharges are well defined on the Hilbert space of
square integrable functions  and generate
$q$-oscillator algebra. As a result, a representation of
$q$-deformed conformal algebra $su_q(1,1)$ is obtained. In Sect.6 we give short
description of further generalizations of the Shabat's potential which
correspond to general $q$-deformed conformal quantum mechanics and
$q$-deformation of (hyper)elliptic potentials.
Sect.7 contains some conclusions.
We would like to stress once more that suggested realizations of
$q$-algebras are continuous (i.e. they are not purely finite-difference
ones) and they are used within the standard physical concepts.

\section{SUSY quantum mechanics}

The simplest $N=2$ SUSY quantum mechanics is fixed by the following
algebraic relations between the Hamiltonian of a system $H$ and
supercharges $Q^{\dag} ,\, Q$ \ci{r19}
\be
\{Q^{\dag},Q\}=H,\quad Q^2=(Q^{\dag})^2=0, \quad [H,Q]=[H,Q^{\dag}]=0.
\lab{e1}
\ee
All operators are supposed to be well defined on the relevant Hilbert
space. Then, independently on explicit realizations the spectrum
is two-fold degenerate and the ground state energy is semipositive,
$E_{vac}\geq 0$.

Let us consider a particle moving in one-dimensional space. Below,
the coordinate $x$ is tacitly assumed to cover the whole line,
$x\in {\it R}$, if it is not explicitly stated that it belongs to some cut.
Standard representation of the algebra \re{e1} contains one free
superpotential $W(x)$ \ci{r20}:
\be
Q=\left(\matrix{0&0\cr A&0\cr}\right), \quad
Q^{\dag}=\left(\matrix{0&A^{\dag}\cr 0&0\cr}\right), \quad
A=(p-iW(x))/\sqrt2,\quad [x,p]=i,
\lab{e2}
\ee
\be
H=\left(\matrix{H_-&0\cr 0&H_+\cr}\right)=
\left(\matrix{A^{\dag} A&0\cr 0&A A^{\dag}\cr}\right)
=\fac12(p^2+W^2(x)-W^\prime(x)\sigma_3),
\lab{e3}
\ee
$$W^\prime(x)\equiv {d\over dx} W(x), \qquad
\sigma_3=\left(\matrix{1&0\cr 0&-1\cr}\right).$$
It describes a particle with two-dimensional internal space the basis
vectors of which can be identified with the spin "up" and "down" states.

The subhamiltonians $H_\pm$ are isospectral as a result of the
intertwining relations
\be
H_- A^{\dag}=A^{\dag} H_+,\qquad A H_-=H_+ A.
\lab{e4}
\ee
The only possible difference concerns the lowest level. Note that the
choice $W(x)=x$ corresponds to the harmonic oscillator problem and
then $A^{\dag},\, A$ coincide with the bosonic creation and
annihilation operators $a^{\dag},\,a$ which satisfy the algebra
\be
[a,a^{\dag}]=1,\qquad [N,a^{\dag}]=a^{\dag},\qquad [N,a]=-a,
\lab{e5}
\ee
where $N$ is the number operator, $N=a^{\dag} a$. This, and another
particular choice, $W(x)=\lambda/x$, correspond to the conformally
invariant dynamics \ci{r21}.

\section{$q$-Deformed SUSY quantum mechanics}

Now we shall introduce the tools needed for the quantum algebraic
deformation of the above construction. Let $T_q$ be smooth $q$-scaling
operator defined on the continuous functions
\be
T_q f(x)=f(qx),
\lab{e6}
\ee
where $q$ is a real non-negative parameter. Evident properties of this
operator are listed below
$$ T_q f(x)g(x)=[T_q f(x)][T_q g(x)],\qquad
T_q {d\over dx}=q^{-1}{d\over dx} T_q, $$
\be
T_q T_p=T_{qp},\qquad T^{-1}_q=T_{q^{-1}},\qquad T_1=1.
\lab{e7}
\ee
On the Hilbert space of square integrable functions ${\cal L}_2$ one has
\be
\int_{-\infty}^{\infty} \phi^*(x)\psi(qx)dx=
q^{-1}\int_{-\infty}^{\infty} \phi^*(q^{-1}x)\psi(x)dx,
\lab{e8}
\ee
where from the hermitian conjugate of $T_q$ can be found
\be
T_q^{\dag}=q^{-1} T_q^{-1},\qquad \quad  (T_q^{\dag})^{\dag}=T_q.
\lab{e9}
\ee
As a result, $\sqrt{\, q}\, T_q$ is a unitary operator.
Because we take wave functions to be infinitely differentiable,
an explicit realization of $T_q$ is provided by the operator
\be
T_q=e^{\ln q\, x\,d/dx}=q^{x\,d/dx}.
\lab{e10}
\ee
Expanding \re{e10} into the formal series and using integration by parts
one can prove relations \re{e9} on the infinite line and semiline $[0,\infty]$.
A special care should be taken for finite cut considerations since $T_q$
moves boundary point(s).

Let us define the $q$-deformed factorization operators
\be
A^{\dag}={1\over \sqrt2}\, (p+iW(x))\,T_q, \qquad
A={q^{-1}\over \sqrt2}\, T_q^{-1} (p-iW(x)),
\lab{e11}
\ee
where $W(x)$ is arbitrary function and for convinience we use
the same notations as in the undeformed case \re{e3}. $A$ and $A^{\dag}$
are hermitian conjugates of each other on ${\cal L}_2$. Now one has
\ba
A^{\dag} A&=&\fac12 q^{-1} (p^2+W^2(x)-W^\prime(x))\equiv q^{-1} H_-,
\lab{e12}   \\
\smallskip
A\, A^{\dag}&=&\fac12q^{-1} T_q^{-1}(p^2+W^2(x)+ W^\prime(x))T_q
\nonumber \\
\smallskip
&=&\fac12q\,(p^2+q^{-2}W^2(q^{-1}x)+ q^{-1}W^\prime (q^{-1}x))
\equiv q H_+.
\lab{e13}
\ea
We define $q$-deformed SUSY Hamiltonian and supercharges to be
\be
H=\left(\matrix{H_-&0\cr 0&H_+\cr}\right)
=\left(\matrix{qA^{\dag} A&0\cr 0&q^{-1}A A^{\dag}\cr}\right),\qquad
Q=\left(\matrix{0&0\cr A&0\cr}\right),\quad
Q^{\dag}=\left(\matrix{0&A^{\dag}\cr 0&0\cr}\right).
\lab{e14}
\ee
These operators satisfy the following $q$-deformed version
of the $N=2$ SUSY algebra
\be
\{Q^{\dag},Q\}_q= H, \quad \{Q,Q\}_q=\{Q^{\dag},Q^{\dag}\}_q=0,\quad
[H,Q]_q=[Q^{\dag}, H]_q=0,
\lab{e15}
\ee
where we introduced $q$-brackets
\be
[X,Y]_q\equiv qXY-q^{-1}YX,\qquad [Y,X]_q=-[X,Y]_{q^{-1}},
\lab{e16}
\ee
\be
\{X,Y\}_q\equiv qXY+q^{-1}YX,\qquad \{Y,X\}_q=\{X,Y\}_{q^{-1}}.
\lab{e17}
\ee
Note that the supercharges are not conserved because they do not commute
with the Hamiltonian (in this respect our algebra principally
differs from the construction of Ref.\ci{r11a}). An interesting
property of the algebra \re{e15} is that it shares with \re{e1}
the semipositiveness of the ground state energy which follows from the
observation that $Q^{\dag},\, Q$ and the operator $q^{-\sigma_3} H$
satisfy ordinary SUSY algebra \re{e1}. Evidently,
in the limit $q\to 1$ one recovers conventional SUSY quantum mechanics.

For the subhamiltonians $H_\pm$ the intertwining relations look as
follows
\be
H_- A^{\dag}=q^2 A^{\dag} H_+,\qquad A H_-=q^2 H_+ A.
\lab{e18}
\ee
Hence, $H_\pm$ are not isospectral but rather $q$-isospectral,
i.e. the spectrum of $H_-$ can be obtained from the spectrum of
$H_+$ just by the $q^2$-factor scaling:
$$
H_+\, \psi^{(+)}=E^{(+)}\psi^{(+)}, \qquad
H_-\, \psi^{(-)}=E^{(-)}\psi^{(-)},
$$
\be
E^{(-)}=q^2\, E^{(+)}, \qquad
\psi^{(-)}\propto A^{\dag} \psi^{(+)}, \quad
\psi^{(+)}\propto A\, \psi^{(-)}.
\lab{e18a}
\ee
Possible exception concerns only the lowest level in the same spirit
as it was in the undeformed SUSY quantum mechanics. If $A^{\dag}, A$
do not have zero modes then there is one-to-one correspondence between
the spectra. We name this situation as a spontaneously broken $q$-SUSY
because for it $E_{vac}>0$. If $A$ (or, $A^{\dag}$) has zero mode
then $q$-SUSY is exact, $E_{vac}=0$, and $H_+$ (or, $H_-$) has one level
less than its superpartner $H_-$ (or, $H_+$).

As a simplest physical example let us consider the case $W(x)=qx$. The
Hamiltonian takes the form
\ba
H&=&\fac12 p^2+\fac14 (q^2+q^{-2})x^2 +\fac14 (q^{-1}-q)+
\fac14 ((q^2-q^{-2})x^2-q-q^{-1})\sigma_3  \nonumber \\
\smallskip
&=& \fac12 p^2 +\fac12 q^{2\sigma_3} x^2 -\fac12 q^{\sigma_3} \sigma_3,
\lab{e19}
\ea
and describes a spin-1/2 particle in the harmonic potential and
related magnetic field along the third axis.
The physical meaning of the deformation parameter $q$ is analogous
to that in the XXZ-model \ci{r3} -- it is a specific interaction constant
in the standard physical sense. This model has exact $q$-SUSY
and if $q^2$ is a rational number then
the spectrum exhibits accidental degeneracies.

\section{General deformation of superpartner Hamiltonians}

Described above $q$-deformation of the SUSY quantum mechanics is by
no means unique. If one chooses in the formulas \re{e11} $T_q$ to be not
$q$-scaling operator but, instead, the shift operator
\be
T_q f(x)=f(x+q), \qquad T_q=e^{q\, d/dx},
\lab{e20}
\ee
then SUSY algebra will not be deformed at all. The superpartner
Hamiltonians will be isospectral and the presence of
$T_q$-operator results in the very simple deformation of old
superpartner potential $U_+(x)\to U_+(x-q)$ (kinetic
term is invariant). Evidently
such deformation does not change the spectrum of $U_+(x)$ and
that is why SUSY algebra remains intact. Nevertheless it
creates new physically relevant SUSY quantum mechanical models. The crucial
point in generating of them was the implication of essentially
infinite order differential operators as the intertwining operators.

A more general $T_q$ is given by the shift operator
in arbitrary coordinate system
\be
T_q f(z(x))= f(z(x)+q), \qquad
T_q=e^{q\,d/dz(x)},\quad {d\over dz}={1\over z^\prime(x)}\,
{d\over dx}.
\lab{e21}
\ee
The effects of choices $z=\ln x$ and $z=x$ were already discussed above.
In general, operator $T_q$ will not preserve the form of
kinetic term in $H_+$-Hamiltonian. Physically, such change
would correspond to the transition from motion of a particle
on flat space to the curved space dynamics. Below we shall
assume the definition \re{e6} but full affine transformation on the line
$$T_q f(x)=f(qx+a)$$
may be used in all formulas without changes.

An interesting question is whether inversion transformation can be joined to
the affine part so that a complete $SL(2)$ group element $z\to (az+b)/(cz+d)$
will enter the formalism in a meaningful way?
Application of the described construction to the higher
dimensional problems is not so straightforward. If variables
separate (spherically symmetric or other special potentials) then it
may work in a parallel with the non-deformed models. In the many-body case
one can perform independent affine transformations for each of the
superselected by fermionic number subhamiltonians and thus to "deform"
these SUSY models as well.

\section{$q$-Deformed conformal quantum mechanics}

Particular  form of the $su(1,1)$ algebra generators can be realized via
the harmonic oscillator creation and annihilation operators \re{e5}
\be
K_+=\fac12 (a^{\dag})^2,\qquad K_-=\fac12 a^2,\qquad
K_0=\fac12 (N+\fac12),
\lab{e22}
\ee
\be
[K_0, K_\pm]=\pm \,K_\pm,\qquad [K_+,K_-]=-2K_0.
\lab{e23}
\ee
This means that harmonic potential has $su(1,1)$  as the
dynamical symmetry algebra, physical states being split into two
irreducible representaions according to their parity.
Let us show that the potential introduced in Ref.\ci{r18}
obeys the quantum conformal symmetry algebra $su_q(1,1)$ in complete
parallel with \re{e22},\re{e23}.

First, we shall rederive this potential within $q$-SUSY physical
situation. Let us consider the Hamiltonian
of a spin-1/2 particle in an external potential $\fac12 U(x)$ and
a magnetic field $\fac12 B(x)$ along the third axis
\be
H=\fac12 (p^2 + U(x) + B(x) \sigma_3)
\lab{e24}
\ee
and impose two conditions: we take magnetic field to be homogeneous
\be
B=-\beta^2 q^{-2} = constant
\lab{e25}
\ee
and require the presence of $q$-SUSY \re{e15}. Equating \re{e24}
and \re{e14} we arrive at the potential
\be
U(x)=W^2(x)-W^{\prime}(x) + \beta^2 q^{-2},
\lab{e26}
\ee
where $W(x)$ satisfies the following mixed finite-difference and
differential equation
\be
W^\prime(x)+qW^\prime (qx)+W^2(x)-q^2 W^2(qx)=2\beta^2.
\lab{e27}
\ee
This is the condition of a self-similarity \ci{r18} which bootstraps
the potential in different points (in Ref.\ci{r17}
$\beta^2= \gamma^2 (1+q^2)/2$ parametrization was used).
Smooth solution of \re{e27} for symmetric potentials
$U(-x)=U(x)$ is given by the following power series
\be
W(x)=\sum_{i=1}^{\infty} c_i\, x^{2i-1}, \qquad
c_i={q^{2i}-1\over q^{2i}+1}{1\over 2i-1}\sum_{m=1}^{i-1}c_{i-m}c_m, \quad
c_1= {2\beta^2\over 1+q^2}.
\lab{e27a}
\ee
In different limits of the parameters several well known
exactly solvable problems arise: 1) Rosen-Morse -- at $q\to 0;\;$
2) P\"oschl-Teller -- at $\beta\propto q\to \infty ;\;$ 3) harmonic
potential -- at $q\to 1;\;$
4) $1/x^2$-potential -- at $q\to 0$ and $\beta \to 0$.
However, strictly speaking for all these limits to be valid one
has to prove their smoothness, e.g., for 4) there may be solutions
for which two limiting procedures do not commute, etc.
Note also that for the case 2) the coordinate range should be
restricted to finite cut because of the presence of singularities.
Infinite soliton solution of Shabat
corresponds to the range $0<q<1$ at fixed $\beta$.
If $q\ne 0, 1, \infty,$ there is no analytical expression for $W(x)$ but
some general properties of this function may be found along the analysis of
Ref.\ci{r18}.

The spectrum can be derived by pure algebraic means.
We already know that the spectra of $H_\pm$ subhamiltonians are
related via the $q^2$-scaling
\be
E^{(-)}_{n+1}=q^2 E^{(+)}_n,
\lab{e27b}
\ee
where the number $n$ numerates levels from below for both spectra.
Because $q$-SUSY is exact in this model the lowest level of $H_-$
corresponds to the first excited state of $H_+$. But due to the restriction
\re{e25} the spectra differ only by a constant,
\be
E^{(-)}_n=E^{(+)}_n -\beta^2 q^{-2},
\lab{e27c}
\ee
Conditions \re{e27b} and \re{e27c} give us the spectrum of $H$
\be
E_{n,m}=\beta^2 \, {q^{-2m}-q^{2n}\over 1-q^2},
\qquad   m=0,1;\; n=0,1,\dots,\infty .
\lab{e28}
\ee
At $q<1$ there are two finite accumulation points, i.e. \re{e28} looks
similar to two-band spectrum. At $q > 1$ energy eigenvalues seem to
grow exponentially to the infinity but there is a catch which does not
allow to identify \re{e28} in this case with real physical spectrum.
In Ref.\ci{r18} it was proven that for $0<q<1$ the superpotential is smooth
and positive at $x=+\infty$. But then
$\psi_0^{(-)}(x)=\exp (-\int^x W(y)dy)$ is a
normalizable wave function defining the ground state of $H_-$- subhamiltonian
and all other states are generated from it without  violation of the
normalizability condition. Therefore relation \re{e28} at $0<q<1$
defines real physical spectrum.

At $q>1$ the series defining $W(x)$ converges only on a finite
interval $|x|<r<\infty$. From inequalities
$$\rho^2\equiv {q^2-1\over q^2+1}\,
< \, {q^{2i}-1\over q^{2i}+1}\,<\, 1, \qquad i>1$$
we have $0<c_i^{(1)}< c_i<c_i^{(2)}$, where $c_i^{(1,2)}$ are defined
by the rule \re{e27a} when $q$-factor on the right hand side is
replaced by $\rho^2$ and $1$ respectively
$(c_1^{(1,2)}=c_1)$. As a result, $1<2\sqrt{c_1} r/\pi< \rho^{-1}$,
which  means that $W(x)$ is smooth only on a cut at the ends of which
it has some singularities. From the basic relation \re{e27} it follows that
these are simple poles with negative unit residues. In fact there should be an
infinite number of simple "primary" and "secondary" poles.
The former ones are characterized by negative unit residues and
location points $x_m$ tending to
$\pi (m+1/2)/\sqrt{c_1},\, m\in Z,$ at $q\to\infty$ ($c_1$ is fixed).
"Secondary" poles are situated at $x=q^n x_m,\, n\in Z^+,$ with
corresponding residues defined by some algebraic equations.
Unfortunately, general analytical structure of the function $W(x)$ is
not known yet, presented above hypothesis needs rigorous proof with exact
identification of all singularities and this is quite challenging problem.

On the other hand, existence of singularities in superpotential does not
allow to take
formal consequences of SUSY as granted. Namely, isospectrality (or,
$q$-isospectrality) of $H_+$ and $H_-$ for the whole line
problem is broken at this point. Hence one is forced
to consider Shr\"odinger operator \re{e24} on a cut
$[-r,r]$ with boundary conditions $\psi_n(\pm r)=0$.
Pole character of $W(x)$ singularities leads to $\psi_0^{(-)}(\pm r)=0$,
i.e. $\psi_0^{(-)}$ is true ground state of $H_-$. It also garantees
that $U_-(x)$ is finite on the physical boundaries, $U_-(\pm r)<\infty$.
Note, however, that the spectrum $E_n$ for such type of problems
can not grow faster than $n^2$ at $n\to \infty$ \ci{r12a}
in apparent contradiction with \re{e28}. This discrepancy is
resolved by observation that action of $T_q$-operator creates
singularities inside the interval $[-r,r]$ so that $U_+(x)$ and $q^2U_+(qx)$
are not isospectral potentials (in ordinary sense) as it was at $q<1$.
Hence, the $q>1$ case
of \re{e28} does not correspond to real physical spectrum of the model.

The number of deformations of a given function is not limited.
The crucial property preserved by the presented above $q$-curling
is the property of exact solvability of "undeformed" Rosen-Morse,
harmonic oscillator, and P\"oschl-Teller potentials.
It is well known that potentials at infinitely small and exact
zero values of a parameter may obey completely different spectra.
In our case, deformation with $q<1$ converts one-level Rosen-Morse problem
into the infinite-level one with
exponentially small energy eigenvalues.
Whether one gets exactly solvable potential
at $q>1$ is an open question but this is quite plausible because at $q=\infty$
a problem with known spectrum arises.

Derivation of the dynamical symmetry algebra is not difficult.
To find that we rewrite relations \re{e12}, \re{e13} for the
superpotential \re{e27}
\be
A^{\dag}A=q^{-1} H+{\beta^2 q^{-1}\over 1-q^2}, \qquad
A\, A^{\dag}=q\, H+{\beta^2 q^{-1}\over 1-q^2},
\lab{e29}
\ee
where H is the Hamiltonian with purely exponential spectrum
\be
H=\fac12 (p^2+W^2(x)-W^{\prime}(x))- {\beta^2 \over 1-q^2},\qquad
E_n=-{\beta^2 \over 1-q^2}\, q^{2n}.
\lab{e29a}
\ee
Evidently,
\be
AA^{\dag}-q^2 A^{\dag}A=\beta^2 q^{-1}.
\lab{e30}
\ee
Normalization of the right hand side of \re{e30} to unity results in the
first relation entering the definition of $q$-deformed Heisenberg-Weyl
algebra.

The shifted Hamiltonian \re{e29a} and $A^{\dag},\, A$ operators satisfy
braid-type commutation relations
$$[A^{\dag},H]_q=[H,A]_q=0,$$
or,
\be
H\, A^{\dag}=q^2A^{\dag} H,\qquad A\, H=q^2H\, A.
\lab{e30a}
\ee
Energy eigenfunctions $| n\rangle $ can be uniquely determined
from the ladder operators action
\be
A^{\dag}|n\rangle =\beta q^{-1/2}\sqrt{\,{1-q^{2(n+1)}\over 1-q^2}}\,
|n+1\rangle ,\qquad
A\, |n\rangle =\beta q^{-1/2}\sqrt{\, {1-q^{2n}\over 1-q^2}}\,
|n-1\rangle .
\lab{e30b}
\ee

It is convinient to introduce the formal number operator
\be
N={\ln [(q^2-1) H/\beta^2]\over \ln q^2},\qquad
N\, |n\rangle =n |n\rangle,
\lab{e31}
\ee
which is defined only on the eigenstates of the Hamiltonian.
Now one can check that operators
\be
a_q={q\over \beta}\,A\, q^{-N/2},\qquad
a^{\dag}_q={q\over \beta}\, q^{-N/2} A^{\dag}
\lab{e32}
\ee
satisfy original $q$-deformed harmonic oscillator algebra
of Biedenharn and Macfarlane \ci{r4,r5}
\be
a_q a^{\dag}_q - q a^{\dag}_q a_q=q^{-N},\quad
[N,a^{\dag}_q]=a^{\dag}_q,\quad [N,a_q]=-a_q.
\lab{e33}
\ee
The quantum conformal algebra $su_q(1,1)$ is realized as follows,
$$K^+={1\over q+q^{-1}}(a_q^{\dag})^2,\qquad
K^-=(K^+)^{\dag},
\qquad K_0=\fac12 (N+\fac12), $$
\be
[K_0, K^{\pm}]=\pm K^{\pm}, \qquad [K^+,K^-]=-
{q^{4K_0}-q^{-4K_0}\over q^2-q^{-2}}.
\lab{e22a}
\ee
Since $H\propto q^{4K_0}$, the dynamical symmetry algebra of the model
is $su_q(1,1)$.
Generators $K^{\pm}$ are parity invariant and therefore even and odd wave
functions belong to different irreducible representations of this algebra.
We conclude that quantum algebras have useful applications even within
the continuous dynamics described by ordinary differential equations.
A different approach to $q$-deformation of conformal quantum mechanics on the
basis of pure finite difference realizations was suggested in Ref.\ci{r21a}.

Let us compare presented model  with the
construction of Ref.\ci{r23}. Kalnins, Levine, and Miller
called as the conformal symmetry generator any differential operator
$L(t)$ which maps solutions of the
time-dependent Schr\"odinger equation to the solutions, i.e. which
satisfies the relation
\be
i\, {\partial \over \partial t}\,  L-[H,L]=
R\, (i\, {\partial \over \partial t}-H),
\lab{e35}
\ee
where $R$ is some operator. On the shell of Schr\" odinger equation
solutions $L(t)$ is conserved and all higher
powers of space derivative, entering the definition of $L(t)$,
can be replaced by the powers of $\partial /\partial t$ and
linear in $\partial /\partial x$ term. But any analytical
function of $\partial /\partial t$ is replaced by the
function of energy when applied to stationary states.
This trick allows
to simulate any infinite order differential operator by the one
linear in space derivative and to prove that a solution with energy
$E$ can always be mapped to the not-necessarily normalizable solution with
the energy $E+f(E)$ where $f(E)$ is arbitrary analytical function.
"On-shell" raising and lowering operators always can be
found if one knows the basis solutions of the
Schr\"odinger equation but sometimes it is easier
to find symmetry generators and use them in search of the spectrum.
In our construction we have "off-shell" symmetry generators, which
map physical states onto each other and
satisfy quantum algebraic relations in the rigorous operator sense.
In this respect our results are complimentary to those of the
Ref.\ci{r23}.

It is clear that affine transformations provide a particular example of
possible potential deformations leading just to scaling of spectra.
In general one can try to find a map
of a given potential with spectrum $E_n$ to a particular related potential
with the spectrum $f(E_n)$ for any analytical function $f(E)$.
A problem of arbitrary non-linear deformation of Lie algebras
was treated in Ref.\ci{r11b} using the symbols of operators
which were not well defined on proper Hilbert space.
Certainly, the method of Ref.\ci{r23} should be
helpful in the analysis of this interesting problem in a more rigorous
fashion and the model presented above shows that sometimes
an "off-shell" realization of symmetry generators can be found.

\section{Factorization method and new potentials}

SUSY quantum mechanics is related to the factorization method of
solving of Schr\"odinger equation [27-29]. Within the latter
approach one has to find solutions of the following nonlinear chain
of coupled differential equations for superpotentials $W_j(x)$
\be
W_j^\prime+W_{j+1}^\prime+W_j^2-W_{j+1}^2=k_{j+1}\equiv \lambda_{j+1}-
\lambda_j, \quad j=0,1,2\dots
\lab{e37}
\ee
where $k_j, \lambda_j$ are some constants.
The Hamiltonians associated to \re{e37} are
\be
2H_j=p^2+U_j(x)=p^2+W^2_j(x)-W^\prime_j(x)+\lambda_j,
\lab{e38}
\ee
$$U_0(x)=W_0^2-W_0^\prime+\lambda_0,\qquad U_{j+1}(x)=U_j(x)+2W_j^\prime (x),$$
where $\lambda_0$ is an  arbitrary energy shift parameter.

SUSY Hamiltonians are obtained by unification of any
two successive pairs $H_j, H_{j+1}$ in a diagonal $2\times 2$ matrix.
Analogous construction for a piece of the chain \re{e38}
with more entries was called an order $N$ parasupersymmetric quantum
mechanics \ci{r25,r25a}. In the latter case relations \re{e37} naturally arise
as the diagonality conditions of a general $(N+1)\times (N+1)$-dimensional
parasupersymmetric Hamiltonian.

If $W_j(x)$'s do not have severe singularities
then the spectra of two operators from \re{e38}
may differ only by a finite number
of lowest levels. Under the additional condition that the functions
\be
\psi_{0}^{(j)}(x)=e^{ -\int^x W_j(y)dy}
\lab{e39}
\ee
are square normalizable one finds the spectrum
\be
H_j\,\psi^{(j)}_n(x)=E_n^{(j)}\,\psi^{(0)}_n(x),
\qquad E_n^{(j)}=\fac12 \lambda_{j+n},
\lab{e40}
\ee
where subscript $n$ numerates levels from below.
In this case \re{e39} represents ground state wave function
of $H_j$ from which one can determine lowest excited states of
$H_{j^\prime},\, j^\prime<j$,
\be
\psi^{(j)}_n(x)\propto (p+iW_j)(p+iW_{j+1})\dots
(p+iW_{j+n-1})\, \psi_{0}^{(j+n)}.
\lab{e41}
\ee
Any exactly solvable discrete spectrum problem can be represented in the
form \re{e37}-\re{e41}. Sometimes it is easier to solve Schr\"odinger
equation by direct construction of the chain of associated Hamiltonians
\re{e38}.
If $U_0(x)$ has only $N$ bound states then there does not exist $W_N(x)$
making $\psi^{(N)}_{0}$ normalizable. If $W_N(x)=0$, then $H_j\, (j<N)$
has exactly $N-j$ levels, the potential $U_j(x)$ is reflectionless and
corresponds to $(N-j)$-soliton solution of the KdV-equation.

In order to solve evidently underdetermined system \re{e37} one has to
impose some closure conditions. At this stage it is an art of a researcher
to find such an Ansatz which allows to generate infinite number of
$W_j$ and $k_j$ from fewer entries. Most of old known examples
are generated by the choice $W_j(x)=a(x)j +b(x) +c(x)/j$ where $a,b,c$ are
some functions determined from the recurrence relations \ci{r26,r27}
(see also \ci{r18}).
New look on the equations \re{e37} was expressed in Ref.\ci{r29}. It was
suggested to consider that chain as some infinite dimensional dynamical
system and to analyze
general constraints reducing it to the finite-dimensional integrable cases.
In particular, it was
shown that very simple periodic closure conditions
\be
W_{j+N}(x)=W_j(x),\qquad \lambda_{j+N}=\lambda_j,
\lab{e42}
\ee
for $N$ odd lead to all known hyperelliptic potentials describing finite-gap
spectra (i.e. those with finite number of permitted bands).
In this case parameters $\lambda_j$ do not, of cause,
coincide with the spectrum. The first non-trivial example appears
at $N=3$ and corresponds to Lame equation with one finite gap in the
spectrum. Equivalently one can consider arising Schrodinger equation
in the Weierstrass form (then periodic potential has singular points
where wave functions are required to be equal to zero) and
again parameters $\lambda_j$ do not coincide with (purely discrete) spectrum.
Note that in the analysis of parasupersymmetric models
\ci{r25,r25a} constants $k_j$ were naturally treated as arbitrary parameters
only occasionally giving the energy levels.

The self-similar potential of Sect.5 was found in Ref.\ci{r18}
by the following Ansatz in the chain \re{e37}
\be
W_i(x)=q^i W(q^i x),
\lab{e43}
\ee
which gives a solution provided $W(x)$ satisfies the equation \re{e27}
and constants $k_j$  are related to each other as follows
\be
k_j\propto q^{2j},\quad j\geq 0.
\lab{e44}
\ee
As it was already discussed, the parameters
$\lambda_j\propto q^{2j}$ give the spectrum of problem at $0<q<1$
and therefore closure \re{e43} seems to be completely different from \re{e42}.
However, described above $q$-SUSY quantum mechanics and
subsequent derivation of \re{e43},\re{e44} shows that in fact
\re{e43} is a $q$-deformation of the following closure condition
\be
W_{j+1}(x)=W_j(x), \qquad k_{j+1}=k_j,
\lab{e45}
\ee
which leads to harmonic oscillator potential. Indeed, one may write
\be
W_{j+1}(x)=qW_j(qx), \qquad  k_{j+1}=q^2 k_j
\lab{e46}
\ee
and check that \re{e43}, \re{e44} follow from these conditions.

As it was announced in Ref.\ci{r18a} one can easily generalize
deformation of SUSY quantum mechanical models to the
parasupersymmetric ones. In the particular case defined by
$(N+1)$-member piece of the chain \re{e38} one simply has to act on the
successive Hamiltonians by different
affine transformation group elements. This would lead to multiparameter
deformation of the parasupersymmetric algebraic relations. Following the
consideration of Ref.\ci{r25} one may impose analogous physical restrictions
on the  Hamiltonians and look for the explicit form of  potentials
accepting these constraints. Analyzing such possibilities
the author have found the following general $q$-periodic closure of the
chain \re{e37}
\be
W_{j+N}(x)=qW_j(qx), \qquad k_{j+N}=q^2 k_j.
\lab{e47}
\ee
These conditions describe $q$-deformation of the finite-gap and related
potentials appearing at $q=1$. Let us find a symmetry algebra
behind \re{e47} at $N=2$.

First we write out explicitly the system of arising equations
\ba
W_1^\prime(x) +W_2^\prime(x)+W_1^2(x)-W^2_2(x)=2\alpha, \nonumber \\
W_2^\prime(x) +qW_1^\prime(qx)+W_2^2(x)-q^2W^2_1(qx)=2\beta.
\lab{e48}
\ea
One can check that the operators
\be
K^+=\fac12(p+iW_1)(p+iW_2)\sqrt{q} T_q,\qquad K^-=(K^+)^{\dag}
\lab{e49}
\ee
satisfy the relations
\be
K^+K^-=H(H-\alpha),\qquad K^-K^+=(q^2 H+\beta)(q^2 H+\alpha+\beta),
\lab{e50}
\ee
$$H=\fac12(p^2+W_1^2(x)-W_1^\prime(x)).$$
The operator $H$ obeys the following commutation relations with $K^\pm$
\be
HK^+-q^2K^+H=(\alpha+\beta)K^+,\qquad
K^-H-q^2HK^-=(\alpha+\beta)K^-.
\lab{e51}
\ee
Note that by adding to $H$ of some constant equations \re{e51} may be
rewritten in the form \re{e30a}.

On the basis of \re{e50} one may define various $q$-commutation relations
between $K^+$ and $K^-$. The simplest one would be the following
\be
K^-K^+-q^4K^+K^-=q^2(\alpha(1+q^2)+2\beta )H+\beta(\alpha+\beta).
\lab{e52}
\ee
The formal map onto the relations \re{e22a} is also available. Therefore
relations \re{e51},\re{e52} give a particular form of the "quantization"
of the algebra $su(1,1)$ which is explicitly recovered at $q=1$.

Described $q$-deformation of the conformal quantum mechanics is more general
than that presented in Sect.5. Indeed, various limits of $q$ give the
following solvable cases: 1) a two-level potential corresponding to
two-soliton system appears at
$q=0$; 2) a finite cut analog of two-soliton potential arises at $q\to\infty$;
3) the general conformal potential
comprising both oscillator and $1/x^2$ parts is recovered in the limit
$q\to1$ when $W(x)\propto a/x+bx$. In order to find the spectrum of $H$
at arbitrary $q$ it is neccessary to know general properties of the
superpotential $W_1$. Let us suppose that there exists a solution
for positive $\alpha$ and $\beta$ such that $\exp (-\int^x W_{1,2})$ are
normalizable wave functions. Then the spectrum consists of two
geometric series and by shifting can be represented in the form
\be
E_n=\cases{E_0 q^{2m},& for n=2m \cr\noalign{\vskip2pt}
           E_1 q^{2m},& for n=2m+1 \cr}
\lab{e53}
\ee
with the $E_n<E_{n+1}$ ordering fulfilled.
Even and odd wave
functions fall into independent irreducible representations of $su_q(1,1)$.
A more detailed consideration of potentials and algebraic structures
arising from the
$q$-periodic closure of the chain \re{e37} will be
given elsewhere.

\section{Conclusions}

To conclude, we described a deformation
of the  SUSY quantum mechanics on the basis of affine transformations.
The main feature of the construction
is that superpartner Hamiltonians satisfy non-trivial braid-type
intertwining relations which remove degeneracies of the original SUSY
spectra. Obtained formalism naturally leads to the Shabat's self-similar
potential describing slowly decreasing solutions of
the KdV equation.
The latter is shown to have straightforward meaning as a
$q$-deformation of the harmonic oscillator potential. Equivalently,
one may consider it as a deformation of a one-soliton system.
Corresponding raising and
lowering operators satisfy $q$-deformed Heisenberg-Weyl algebra atop of
which a quantum conformal algebra $su_q(1,1)$ can be built.
We also
outlined a generalization of the Shabat's potential on the basis of
$q$-deformation of periodic closure condition and presented $q$-deformation
of general conformal quantum mechanics potentials.

In this paper the parameter $q$ was taken to be real but nothing prevents
from consideration
of complex values as well (this changes only hermicity properties).
The most interesting cases appear when $q$ is
a root of unity \ci{r31}. For example, at $q^3=1$ eq. \re{e27} generates a
potential proportional to the so-called equianharmonic Weierstrass functions.
More complicated
hyperelliptic potentials are generated at higher roots of unity.
The nontrivial Hopf algebra structure of the quantum groups was not considered
because it is not relevant in the context of quantum mechanics of one
particle in one dimension. Perhaps higher dimensional and many body
problems shall elucidate this point. In fact, there seems to be no
principle obstacles for higher dimensional generalizations
although resulting systems may not have direct physical meaning.
Another possibility is that described self-similar systems
may arise from higher dimensional ones after the similarity
reductions.

In order to illustrate various possibilities we
rewrite the simplest self-similarity equation without scaling (i.e. at $q=1$)
but with non-trivial translationary part
\be
W^\prime(x)+W^\prime(x+a)+W^2(x)-W^2(x+a)=constant.
\lab{e54}
\ee
Solutions of this equation provide a realization of the ordinary
undeformed Heisenberg-Weyl algebra.
The full effect of the presence of the parameter $a$ in \re{e54} is
not known to the author but solutions whose absolute values
monotonically increase at
$x\to\pm \infty$ seem to be forbidden.
Note also that in all formulas of SUSY and $q$-SUSY quantum mechanics
superpotential $W(x)$ may be replaced by a hermitian $n\times n$ matrix
function. The equations \re{e27}, \re{e30}, \re{e54} may be equally thought as
being the
matrix ones with the right hand sides proportional to unit matrices.
We end by a speculative conjecture that described machinery may
be useful in seeking for $q$-deformations of the non-linear
integrable evolution equations, like KdV, {\it sin}-Gordon, etc.

\section{Acknowledgments}

The author is indebted to J.LeTourneux, W.Miller, A.Shabat, L.Vinet
for valuable discussions and to Y.S.Kim for kind invitation to present
this paper at the Workshop on Harmonic Oscillators.
This research was supported by the NSERC of Canada.


\bb{40}

\bi{r1} {\it Dynamical Groups and Spectrum Generating Algebras},
        Eds. A.Bohm, Y.Ne'eman, and A.Barut (World Scientific, 1988);
Y.Alhassid, F.G\"ursey, and F.Iachello, Phys.Rev.Lett. {\bf 50}, 873 (1982);
        Ann.Phys. (N.Y.) {\bf 148}, 346 (1983);
        M.Moshinsky, C.Quesne, and G.Loyola, Ann.Phys. (N.Y.) {\bf 198},
        103 (1990).

\bi{r2} V.G.Drinfeld, Quantum Groups, {\it in:} Proc. of the Intern.
      Congress of Mathematicians (Berkeley, 1986) vol.1, p.798;
        M.Jimbo, Lett.Math.Phys. {\bf 10}, 63 (1985); {\bf 11}, 247
        (1986);  N.Yu.Reshetikhin, L.A.Takhtajan, and
        L.D.Faddeev, Algebra i Analiz, {\bf 1}, 178 (1989).

\bi{r3} N.Pasquier and H.Saleur, Nucl.Phys. {\bf B330}, 523 (1990);
M.T.Batchelor, L.Mezinchescu, R.I.Nepomechie and V.Rittenberg, J.Phys.
{\bf A23}, L141 (1990).

\bi{r4} L.C.Biedenharn, J.Phys. {\bf A22}, L873 (1989).

\bi{r5} A.J.Macfarlane, J.Phys. {\bf A22}, 4581 (1989).

\bi{r6} C.-P. Sen and H.-C.Fu, J.Phys. {\bf A22}, L983 (1989);
        T.Hayashi, Comm.Math.Phys. {\bf 127}, 129 (1990);
        M.Chaichian and P.Kulish, Phys.Lett. {\bf B234}, 72 (1990);
N.M.Atakishiev and S.K.Suslov, Sov.J.Theor.Math.Phys.
 {\bf 85}, 1055 (1990);
        P.P.Kulish and E.V.Damaskinsky, J.Phys. {\bf A23}, L415 (1990);
        R.Floreanini, V.P.Spiridonov, and L.Vinet, Comm.Math.Phys.
        {\bf 137}, 149 (1991);
        D.B.Fairlie and C.Zachos, Phys.Lett. {\bf B256}, 43 (1991);
        E.G.Floratos, J.Phys. {\bf A24}, 4739 (1991);
  R.Floreanini, D.Leites, and L.Vinet, Lett.Math.Phys. {\bf 23}, 127 (1992);
       R.Floreanini and L.Vinet, Lett.Math.Phys. {\bf 23}, 151 (1992).

\bi{r8} R.Floreanini and L.Vinet, Representations of Quantum Algebras
        and $q$-Special Functions, {\it in:} Proc. of the ${\it II^{nd}}$
        Intern. Wigner Symposium, to be published.

\bi{r9} E.G.Floratos and T.N.Tomaras, Phys.Lett. {\bf B251}, 163 (1990).

\bi{r10}J.Wess and B.Zumino, Nucl.Phys. (Proc.Suppl.) {\bf B18}, 302
        (1990);  B.Zumino, Mod.Phys.Lett. {\bf A6}, 1225 (1991);
        U.Carow-Watamura, M.Schlieker, and S.Watamura, Z.Phys.
        {\bf C49}, 439 (1991);
        J.A.Minanhan, Mod.Phys.Lett. {\bf A5}, 2635 (1990);
        L.Baulieu and E.G.Floratos, Phys.Lett. {\bf B258}, 171 (1991).

\bi{r11} P.P.Raychev, R.P.Roussev, and Yu.F.Smirnov, J.Phys. {\bf G16},
         L137 (1990);
         M.Chaichian, D.Ellinas, and P.Kulish, Phys.Rev.Lett. {\bf 65},
         980 (1990);
         R.M.Mir-Kasimov, The Relativistic Oscillator as the Realization
         of the Quantum Group of Dynamical Symmetry, {\it in:} Proc.
         of the Intern. Seminar "Quarks'90", 14-19 May 1990, Telavi,
     USSR. Eds. V.A.Matveev et al (World Scientific, Singapore) p.133.

\bi{r11a} M.Chaichian, P.Kulish, and J.Lukierski, Phys.Lett.
        {\bf B262}, 43 (1991).

\bi{r11b} A.P.Polychronakos, Mod.Phys.Lett. {\bf A5}, 2325 (1990);
A.T.Filippov, D.Gangopadhyay, and A.P.Isaev, J.Phys. {\bf A24}, L63 (1991);
         M.Ro{\v c}ek, Phys.Lett. {\bf B255}, 554 (1991);
         K.Odaka, T.Kishi, and S.Kamefuchi, J.Phys. {\bf A24}, L591 (1991);
         C.Daskaloyannis, J.Phys. {\bf A24}, L789 (1991);
         C.Daskaloyannis and K.Ypsilantis, preprint THES-TP-91/09, 1991.

\bi{r11c} O.W.Greenberg, Phys.Rev.Lett. {\bf 64}, 705 (1990) and talk
at this workshop;
        R.Mohapatra, Phys.Lett. {\bf B242}, 407 (1990);
        V.P.Spiridonov, Dynamical Parasupersymmetry in Quantum
        Systems, {\it in:} Proc. of the Intern. Seminar "Quarks'90",
        14-19 May 1990, Telavi, USSR. Eds. V.A.Matveev et al (World
        Scientific, Singapore) p.232

\bi{r12} M.M.Nieto, Phys.Lett. {\bf B145}, 208 (1984);
         M.Luban and D.L.Pursey, Phys.Rev. {\bf D33}, 431 (1986).

\bi{r12a} V.A.Marchenko, {\it Sturm-Liouville Operators and Applications}
          (Birkh\"auser Verlag, 1986).

\bi{r15} A.Yu.Ignatiev and V.A.Kuzmin, Yad.Fiz. {\bf 46}, 786 (1987).

\bi{r16a} I.B.Frenkel and N.Yu.Reshetikhin, Quantum Affine Algebras,
Commutative Systems of Difference Equations and Elliptic Solutions to
the Yang-Baxter Equations, {\it in:} Proc. of the $\DGM$ Intern. Conf.
on the Diff. Geom. Methods in Theor. Physics, New York, USA, 3-7
June 1991. Eds. S.Catto and A.Rocha (World Scientific, 1992) p.46.

\bi{r16b} P.P.Kulish, Quantum Groups and Quantum Algebras as Symmetries
of Dynamical Systems, {\it in:} Proc. of the ${\it II^{nd}}$ Intern.
Wigner Symposium, to be published.

\bi{r18} A.Shabat, Inverse Prob. {\bf 8}, 303 (1992).

\bi{r17} V.Spiridonov, Phys.Rev.Lett. {\bf 69}, 398 (1992).

\bi{r19} L.E.Gendenstein and I.V.Krive, Sov.Phys.Usp. {\bf 28}, 645 (1985);
    V.Kosteleck\'y, talk at this workshop.

\bi{r20} E.Witten, Nucl.Phys. {\bf B188}, 513 (1981).

\bi{r18a} V.Spiridonov, Mod.Phys.Lett. {\bf A7}, 1241 (1992).

\bi{r21} V.DeAlfaro, S.Fubini, and G.Furlan, Nuovo Cim. {\bf A34}, 569 (1976).

\bi{r21a} R.Floreanini and L.Vinet, Phys.Lett. {\bf B277}, 442 (1992).

\bi{r23} E.G.Kalnins, R.D.Levine, and W.Miller, Jr., Conformal Symmetries
        and Generalized Recurrences for Heat and Schr\"odinger
        Equations in One Spatial Dimension, {\it in:}
        Mechanics, Analysis and Geometry: 200 Years after Lagrange,
        Ed. M.Francaviglia (Elsevier Science Publishers B.V., 1991) p.237

\bi{r26} L.Infeld and T.E.Hull, Rev.Mod.Phys. {\bf 23}, 21 (1951).

\bi{r27} W.Miller, Jr., {\it Lie Theory and Special Functions}
(Academic Press, 1968).

\bi{r28} L.E.Gendenstein, Pis'ma ZhETF {\bf 38}, 299 (1983).

\bi{r25} V.A.Rubakov and V.P.Spiridonov, Mod. Phys. Lett. {\bf A3}, 1337
         (1988);  V.Spiridonov, Parasupersymmetry in Quantum
         Systems, {\it in:} Proc. of the $\DGM$ Intern. Conf. on the Diff.
         Geom. Methods in Theor. Physics, New York, USA, 3-7 June 1991.
Eds. S.Catto and A.Rocha   (World Scientific, 1992) p.622.

\bi{r25a}  S.Durand, M.Mayrand, V.P.Spiridonov, and L.Vinet, Mod.Phys.Lett.
{\bf A6}, 3163 (1991).

\bi{r29} A.B.Shabat and R.I.Yamilov, Algebra i Analiz, {\bf 2}, 377 (1991).

\bi{r31} A.Shabat and V.Spiridonov, unpublished.

\eb
\end{document}